\documentclass[aps,twocolumn,showpacs]{revtex4}

\usepackage{epsfig}

\begin{document}

\title{Quantum discord in spin-cluster materials}

\author{
Mikhail A. Yurischev
}
\email{yur@itp.ac.ru}
\affiliation{
Institute of Problems of Chemical Physics
of the Russian Academy of Sciences,
142432 Chernogolovka,
Moscow Region,
Russia
}

\begin{abstract}
The total quantum correlation (discord) in Heisenberg dimers
is expressed via the spin-spin correlation function, internal energy,
specific heat or magnetic susceptibility.
This allows one to indirectly measure the discord through
neutron scattering, as well as calorimetric or magnetometric
experiments.
Using the available experimental data, we found the discord for a
number of binuclear Heisenberg substances with both antiferro- and
ferromagnetic interactions.
For the dimerized antiferromagnet copper nitrate
${\rm Cu(NO_3)_2\cdot2.5 H_2O}$,
the three independent experimental methods named above lead
to a discord of approximately 0.2--0.3~bit/dimer at a
temperature of 4~K.
We also determined the temperature behavior of discord for hydrated
and anhydrous copper acetates, as well as for the ferromagnetic
binuclear copper acetate complex
${\rm [Cu_2L(OAc)]\cdot6H_2O}$, where L is a ligand.
\end{abstract}

\medskip

\pacs{03.65.Ud, 03.67.-a, 75.10.Jm, 75.50.Xx}

\maketitle



\section{Introduction}
\label{sec:Intro}
In the last years, it was understood that entanglement does not
exhaust all quantum correlations in a system.
Information correlations measured by entropy \cite{Lind73,AC97,Hol10}
were placed at the forefront.
The total amount of correlations is identified with the 
mutual information $I$.
The total correlations may be divided into two parts: purely classical
correlations $C$ and purely quantum ones $Q$
\cite{HVed01,OZur02,Ved03}.
The quantum excess of correlations, $Q=I-C$, has been called
{\em discord\/} \cite{OZur02}. 

Note that the quantum discord has been evaluated explicitly only
for several families of two-particle states (density
matrices) \cite{Luo08,FWBAC10,ARA10,GA11}. 
 
It is remarkable that discord can exist even in separable (but
mixed) states, i.e. when quantum entanglement is identically equal
to zero.
An example of two-qubit separable state with nonzero discord
$Q=(3/4){\rm log}_2(4/3)\simeq0.311$
was given in Ref.~\cite{DSC08}.
Below, in our paper, we will also discuss similar situations.
As it turns out, ``almost all quantum states have
nonclassical correlations'' \cite{FACCA10}.
Thus, quantum discord is a different measure of quantum correlation
than entanglement.

As shown on the model of deterministic quantum computation with one
pure qubit (DQC1) \cite{DSC08,LBAW08}, quantum discord can lead to
a speedup over classical computation even without containing much
entanglement.
Discord can also detect the quantum phase transitions
\cite{Dil08,Sar09}.
Moreover, recently it has been shown \cite{WTRR10} that in
contrast to the entanglement and other thermodynamical quantities,
discord makes it possible to identify the quantum phase
transition points at {\em finite\/} temperatures $(T>0)$. 

However, in order to utilize in practice the remarkable properties of
quantum discord it is necessary to find methods to measure it 
experimentally.
Our paper concerns this important problem.
For the two-qubit Heisenberg systems, we found the relations
between discord and the ordinary spin-spin correlation
function, as well as between discord and internal energy, specific heat
or magnetic susceptibility.
This allows one to determine the behavior of quantum discord in various 
substances and investigate its properties. 

The remaining part of this paper is organized as follows.
In Section 2, we describe the model and present formulas for
calculating the total, classical, and quantum correlations.
Here a comparison analysis is also made for those correlations and
the spin-spin correlation function and for entanglement.
Section 3 is devoted to expressions of discord in terms of spin-spin
correlation function, internal energy, specific heat, and magnetic
susceptibility which are measured by standard experiments.
These approaches are used on the copper nitrate compound.
In this section we also present the temperature dependencies for the
discord in crystals of copper(II) acetate complexes, which are excellent
examples of dimeric materials.
The results obtained are briefly summarized in Section 4.

\section{Correlations in Heisenberg dimer}
\label{sec:CHD}
The Hamiltonian of a Heisenberg dimer reads
\begin{equation}
   \label{eq:H}
   {\cal H}=-{1\over2}\,J{\vec\sigma}_1{\vec\sigma}_2 , 
\end{equation}
where $J$ is the exchange coupling constant and
${\vec\sigma_i}=(\sigma_i^x,\sigma_i^y,\sigma_i^z)$ the vector of
Pauli matrices at the site $i=1$ or 2. 
The magnetic moment components for the dimer are
\begin{equation}
   \label{eq:M}
   M_\nu={1\over2}\,g_\nu\mu_B(\sigma_1^\nu + \sigma_2^\nu),
   \qquad \nu=x, y, z . 
\end{equation}
Here, $g_\nu$ are the components of the $g$ factor and $\mu_B$ is the
Bohr magneton.

The density matrix of a system in thermal equilibrium has the Gibbs
form
\begin{equation}
   \label{eq:rho}
   \rho={1\over Z}\exp(-{\cal H}/k_BT),
\end{equation}
where $k_B$ is the Boltzmann constant and $Z$ the partition function
which is found from the condition ${\rm Tr}\rho=1$.
Performing necessary calculations for the system (\ref{eq:H}) one
arrives at 
\begin{equation}
   \label{eq:rhoT}
   \rho(T)={1\over4}\,(1+G{\vec\sigma}_1{\vec\sigma}_2)
   ={1\over 4}\left(
      \begin{array}{cccc}
      1+G&&&\\
      &1-G&2G&\\
      &2G&1-G&\\
      &&&1+G
      \end{array}
   \right)
\end{equation}
with
\begin{equation}
   \label{eq:G}
   G(T)=-1+\frac{4}{3+\exp(-2J/k_BT)} .
\end{equation}
It is easy to check that the quantity $G$ equals the spin-spin
correlation functions,
\begin{equation}
   \label{eq:Gss}
   G=\langle\sigma_1^x\sigma_2^x\rangle
    =\langle\sigma_1^y\sigma_2^y\rangle
    =\langle\sigma_1^z\sigma_2^z\rangle , 
\end{equation}
where the brackets denote the statistical average.
Values of $G$ range from $-1$ to zero for the antiferromagnetic
cluster $(J<0)$ and from zero to 1/3 for the ferromagnetic one
$(J>0)$.

The density matrix (\ref{eq:rhoT}) has the form for which the quantum
discord is evaluated exactly \cite{Luo08}:
\begin{equation}
   \label{eq:Q}
   Q=I-C , 
\end{equation}
where the mutual information equals
\begin{equation}
   \label{eq:I}
   I={1\over4}\,[(1-3G)\,{\rm log}_2(1-3G)
    +3(1+G){\rm log}_2(1+G)]
\end{equation}
and the classical part of the total correlations is
\begin{equation}
   \label{eq:C}
   C={1\over2}\,[(1+|G|){\rm log}_2(1+|G|)
    +(1-|G|){\rm log}_2(1-|G|)] .
\end{equation}
Expressions (\ref{eq:G}), (\ref{eq:Q})--(\ref{eq:C}) define so-called
thermal discord \cite{WR10,WLZ10,PB10}.

Another type of quantum correlation in a system is the
entanglement of formation
\begin{eqnarray}
   \label{eq:E}
   E&=&
   -\frac{1+\sqrt{1-\tilde C^2}}{2}\,
   {\rm log}_2\left(\frac{1+\sqrt{1-\tilde C^2}}{2}\right)
   \nonumber\\
   &&
   -\frac{1-\sqrt{1-\tilde C^2}}{2}\,
   {\rm log}_2\left(\frac{1-\sqrt{1-\tilde C^2}}{2}\right) , 
\end{eqnarray}
where concurrence $\tilde C$ for the ferromagnetic dimer completely
vanishes, and for the antiferromagnetic one is \cite{N98,ABV01}
\begin{eqnarray}
   \label{eq:Cd}
   \tilde C(T)=\cases{
   -{1\over2}(1+3G),& $T<T_e$;\cr
   0,& $T\ge T_e$.\cr} 
\end{eqnarray}
Here the temperature $T_e$ is given by equation 
\begin{equation}
   \label{eq:Te}
   \frac{k_B}{|J|}\,T_e={2\over \ln3}\,=1.8204\ldots .
\end{equation}
According to Eqs.~(\ref{eq:Q})--(\ref{eq:Te}), quantum discord
and entanglement are functionally related.

\begin{figure}[t]
\begin{center}
\epsfig{file=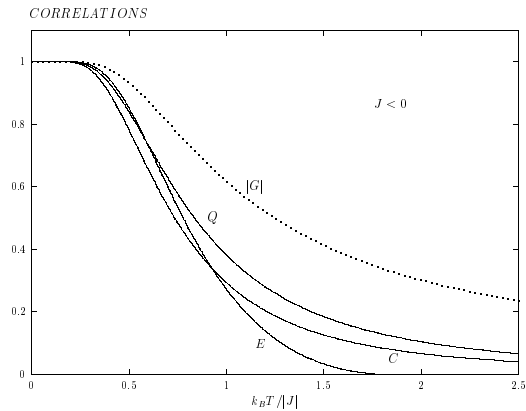,width=6.8cm}
\caption{
The temperature dependencies of correlations $|G|$, $Q$, $C$, and $E$
for the antiferromagnetic Heisenberg dimer.}
\label{fig:afQ}
\end{center}
\end{figure}

In Fig.~\ref{fig:afQ} the temperature dependencies of $|G|$, $E$, $Q$,
and $C$, which describe different correlations in the antiferromagnetic
dimer are shown.
(In an antiferromagnet, the correlator $G\leq0$ therefore we took
its absolute value.) 
It is seen that at $T=0$ all correlations are maximal (saturated)and
equal to one.
The correlations preserve practically the same value with a small
increase of temperature.
This is associated with the existence of a gap in the energy
spectrum of the discussed system.
With a further increase in temperature, all kinds of correlations
monotonically decrease.
By this, the curves pass through the inflection points where a convexity
is changed to concavity.
Functions $|G(T)|$, $Q(T)$, and $C(T)$ are different from zero for
all temperatures $T<\infty$.
When $T\to\infty$, the value of spin-spin correlations tends
to zero as $|G|\sim1/T$, and the information correlations $Q(T)$
and $C(T)$ go to zero more rapidly --- according to the law
$1/T^2$.

From Fig.~\ref{fig:afQ} it can be seen that at low temperatures the
entanglement $E$ is larger than both $Q$ and $C$.
Because $E(T)$ has the sudden disappear point at the temperature $T_e$,
the curve $E(T)$ must intersect both $Q(T)$ and $C(T)$.
Coordinates of intersection points are
\begin{eqnarray}
   \label{eq:QE}
   &&k_BT_{QE}/|J|=0.5880\ldots ,
   \nonumber\\
   &&
   Q(T_{QE})=E(T_{QE})=0.7462\ldots\simeq0.75 
\end{eqnarray}
and
\begin{eqnarray}
   \label{eq:CE}
   &&k_BT_{CE}/|J|=0.9260\ldots ,
   \nonumber\\
   &&
   Q(T_{CE})=E(T_{CE})=0.3390\ldots\, .
\end{eqnarray}
The phenomenon that in some situations the entanglement can be larger
that the total quantum correlations has been pointed out in
Ref.~\cite{Luo08}.
From our calculations, it follows that by $T\neq0$ discord is always
greater than classical correlation, and the latter may be both larger 
and smaller than quantum entanglement.

At the temperature $T_e$, entanglement in the antiferromagnetic dimer
vanishes.
On the other hand, the spin correlation $G(T_e)=-1/3$, and the total
entropy correlations
\begin{equation}
   \label{eq:Ie}
   I_e\equiv I(T_e)=1-{1\over2}\,{\rm log}_23=0.2075\ldots
\end{equation}
and discord
\begin{equation}
   \label{eq:Qe}
   Q_e\equiv Q(T_e)={1\over2}\,{\rm log}_23 - {2\over3}=0.1258\ldots\,.
\end{equation}

Consider now a dimer with the ferromagnetic coupling $J>0$.
At zero temperature, $G=1/3$ and the density matrix (\ref{eq:rhoT}) is
\begin{eqnarray}
   \label{eq:rho0}
   &&\rho_0\equiv\rho(0)
   ={1\over6}\left(
      \begin{array}{cccc}
      2&&&\\
      &1&1&\\
      &1&1&\\
      &&&2
      \end{array}
   \right)
   ={1\over6}[2(|00\rangle\langle00|
   \nonumber\\
   &&
   +|11\rangle\langle11|)
   +(|01\rangle+|10\rangle)(\langle01|+\langle10|)] .
\end{eqnarray}
Here
$\{|00\rangle, |01\rangle, |10\rangle, |11\rangle\}$
is the standard basis for the two-qubit system.
The state (\ref{eq:rho0}) is mixed because $\rho_0^2\neq\rho_0$.
At the same time, the state (\ref{eq:rho0}) is separable (i.e., the
entanglement $E=0$).
Indeed, a partial transposition of $\rho_0$ is
\begin{equation}
   \label{eq:rho0t}
   \rho_0^{t_i}
   ={1\over6}\left(
      \begin{array}{cccc}
      2&.&.&1\\
      .&1&.&.\\
      .&.&1&.\\
      1&.&.&2
      \end{array}
   \right) .
\end{equation}
Eigenvalues of this matrix equal $1, 1, 1$, and 3.
All these eigenvalues are positive.
Consequently, in accordance with the positive partial transpose (PPT)
criterion \cite{Per96,HHH96}, the state (\ref{eq:rho0}) is separable.
Using formulas (\ref{eq:Q})--(\ref{eq:C}) we find that the discord
of a state (\ref{eq:rho0}) equals $Q_0=1/3\simeq0.333$.
This value is larger than $0.311$ for the discord of the example 
from Ref.~\cite{DSC08} mentioned above.

\begin{figure}[t]
\begin{center}
\epsfig{file=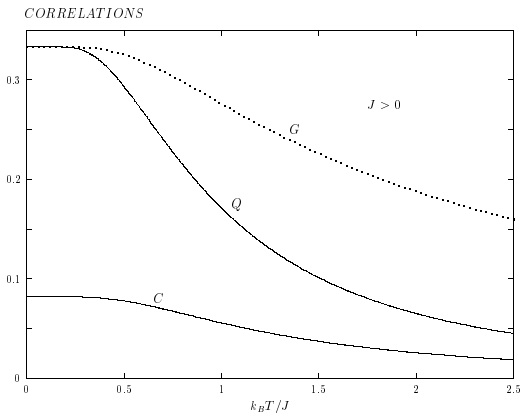,width=6.8cm}
\caption{
Correlations $G$, $Q$, and $C$ vs $k_BT/J$ in the ferromagnetic
Heisenberg dimer.}
\label{fig:fQ}
\end{center}
\end{figure}

Let us look at Fig.~\ref{fig:fQ}.
The figure shows the dependencies of $G$, $Q$, and $C$ versus
temperature for the dimer with the ferromagnetic coupling.
Entanglement is absent in such a system.
At zero temperature, the spin-spin correlation reaches its maximal
value $G=1/3$.
Here quantum discord is also equal to 1/3.
It is interesting that the total classical correlations
are now less than $Q$ and according to (\ref{eq:C}) equal only to
\begin{equation}
   \label{eq:C0}
   C_0\equiv C(0)={5\over3} - {\rm log}_23=0.0817\ldots\, .
\end{equation}
Thus, the ratio of discord to classical correlation here achieves
the value $Q_0/C_0=1/[5-3{\rm log}_23]\approx4.0798$
(c.f. Ref.~\cite{GGZ11}).

At low temperatures all three types of correlations have
quasi-horizontal sections (``pedestals'').
With increasing temperature, the correlations pass through inflection
points and then decay asymptotically to zero.
Curves $G(T)$, $Q(T)$, and $C(T)$ do not at any point intersect with
one another.

\begin{figure}[t]
\begin{center}
\epsfig{file=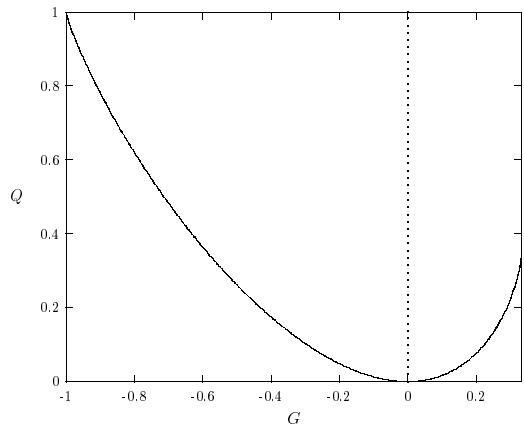,width=6.8cm}
\caption{
Discord as a function of $G$.
Dashed vertical line separates antiferro- and ferromagnetic regions.}
\label{fig:QG}
\end{center}
\end{figure}

Notice the following.
In accordance with Eqs.~(\ref{eq:Q})--(\ref{eq:C}), the quantum discord
$Q$ is a function of $G$ (see Fig.~\ref{fig:QG}). 
This function is monotonic both for the antiferromagnetic and
ferromagnetic dimers.
Moreover, if $G=0$ (absence of ordinary correlations) then also $Q=0$
(absence of any quantum correlations), when $|G|$ is maximum
($=1$ or 1/3) then $Q$ takes the same maximum values.
This allows us to consider spin-spin correlation as a measure of
discord. 
This is similar to that of the concurrence as a measure of
entanglement \cite{HW97,W98}.

\section{Experimental measurements of discord}
\label{sec:Qchi}

As mentioned above, discord has a number of attractive properties.
Unfortunately, the question of how to measure it experimentally is
open.
But in the case of spin dimers, the information correlations $I$, $C$,
and $Q$ can be expressed via the experimentally observed
characteristics of a system.

\subsection{Neutron scattering and discord}
\label{sec:ns}

Inelastic scattering of thermal neutrons is a powerful tool for the
study of low-energy excitations in crystalline transition-metal and
other compounds.
By this, the Fourier components of spin
pair-correlation function
are extracted from the scattering data
(\cite{TNGBT97,HBM05,STL07} and references therein).
Performing the inverse Fourier transformation produces the 
correlation function itself.

The neutron scattering experimental results for the quasi-dimer
antiferromagnetic crystals of deuterated copper(II) nitrate
${\rm Cu(NO_3)_2\cdot2.5 D_2O}$
are presented in Ref.~\cite{XBRA00} (see also \cite{BVZ06}).
(Notice that the deuteration needed for performing the neutron
scattering experiments has no measurable effect on the exchange
coupling \cite{GSF79}.)
In the experiments with the copper nitrate, the neutron scattering
intensity was measured for the temperature range $0.31<T<7.66$~K.
The temperature $T_e$ is equal to about 5~K \cite{BVZ06}.
At $T=4$~K, the spin-spin correlation has the value $G=-0.54(9)$
\cite{comment}.
Using Eqs.~(\ref{eq:Q})--(\ref{eq:C}) we estimate the discord as
$Q=0.3(1)$.

Note, in macroscopic systems, molar discord is ${\cal Q}=N_AQ$
($N_A$ is the Avogadro number).
However, we will normalize discord per a dimer.

\subsection{Internal energy and specific heat}
\label{sec:sh}

The internal energy per a mole of dimers (\ref{eq:H}) is given as 
\begin{equation}
   \label{eq:u}
   u(T)=-\frac{3RJ}{2k_B}\,G(T) , 
\end{equation}
where $R=k_BN_A$ is the universal gas constant.
In turn, the energy equals
\begin{equation}
   \label{eq:ucm}
   u(T)=u_0 + \int\limits_0^Tc_m(T)dT ,
\end{equation}
where $c_m(T)$ is the magnetic part of specific heat
(the part after subtraction of lattice contribution from the
total heat capacity).
The integration constant $u_0$ in Eq.~(\ref{eq:ucm}) can be restored
from the condition $u(\infty)=0$, i.e.,
\begin{equation}
   \label{eq:u0}
   u_0=u(0)=-\int\limits_0^\infty c_m(T)dT . 
\end{equation}
Therefore, we can get the correlation function $G$, from
calorimetric measurements, and then find the discord using the
expression of $Q$ via $G$.

On the other hand, the magnetic specific heat per mole of
{\em dimers\/} is given as \cite{Car86}
\begin{equation}
   \label{eq:cmT}
   c_m(T)=12R\left(\frac{J}{k_BT}\right)^2\frac{
   \exp(2J/k_BT)}{1+3\exp(2J/k_BT)} .
\end{equation}
This function exhibits a maximum
(a Schottky-like anomaly).
For the ferromagnetic coupling, $J>0$, its coordinates are
\begin{equation}
   \label{eq:max-cf}
   k_BT_{\rm max}/J=0.9259\ldots ,
   \qquad c_m^{\rm max}/R=0.1663\ldots , 
\end{equation}
and for the antiferromagnetic one, $J<0$,
\begin{equation}
   \label{eq:max-c-af}
   k_BT_{\rm max}/|J|=0.7029\ldots ,
   \qquad c_m^{\rm max}/R=1.0234\ldots\, . 
\end{equation}

\begin{figure}[t]
\begin{center}
\epsfig{file=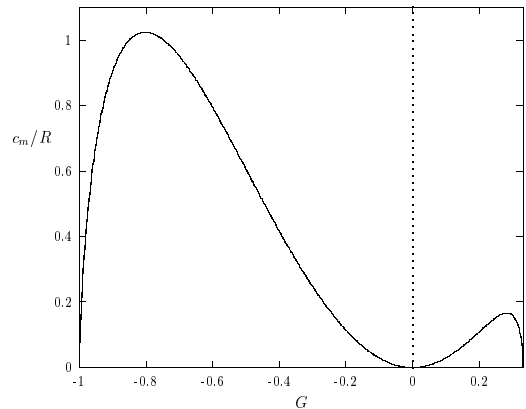,width=6.8cm}
\caption{
Relation of specific heat $c_m/R$ and correlation $G$.
Dashed vertical line separates antiferro- and ferromagnetic regions.
Points $G=-1$ and $G=1/3$ correspond to the temperature $T=0$ and
$G=0$ to the $T=\infty$.}
\label{fig:cG}
\end{center}
\end{figure}

Using Eq.~(\ref{eq:G}) we can rewrite expression for the specific heat
of Heisenberg dimers in the following forms: 
\begin{equation}
   \label{eq:cmTG}
   c_m(T)=\frac{3R}{4}\left(\frac{J}{k_BT}\right)^2(1+G(T))(1-3G(T)) 
\end{equation}
and, especially remarkably,
\begin{equation}
   \label{eq:cmG}
   c_m/R=\frac{3}{16}\,(1+G)(1-3G)\ln^2\left(\frac{1+G}{1-3G}\right) . 
\end{equation}
These relations can be used to extract the correlation $G$
directly from specific-heat measurements.
The behavior of $c_m/R$ versus $G$ is shown in Fig.~\ref{fig:cG}.

Consider as an example copper nitrate
${\rm Cu(NO_3)_2\cdot2.5 H_2O}$.
Its specific heat was measured for the temperature range
$0.5-4.2$~K \cite{FR68} (see also \cite{BFKMB83} and references
therein). 
At temperatures below $0.5$~K, specific heat is zero.
Near $1.82$~K it has the maximum value of $2\times0.51 R$.
At higher temperatures, specific heat decreases obeying at $T\ge4$~K
the asymptotical law \cite{FR68}
\begin{equation}
   \label{eq:cmT2}
   c_m/2R=3.3/T^2 .
\end{equation}
The value of the exchange constant has been estimated as
$2J/k_B=-5.18$~K \cite{FR68}.

We have numerically integrated the available experimental data
from zero to 4~K.
Taking $u(0)/R=3J/2k_B=-3.885$~K
we determined that at $T=4$~K the internal energy is
$u(4)/R=-1.63$~K and hence $G(4)\simeq-0.42$ [see Eq.~(\ref{eq:u})].
Consequently, according to the calorimetric data, the discord
in the given binuclear cluster material at the temperature of 4~K
equals $Q=0.19$.
This value can be considered to be reasonably in agreement with
the estimate obtained from the neutron scattering measurements.  

Note that the integration of function (\ref{eq:cmT2}) from 4~K to
infinity leads to $u(4)/R=-1.65$~K.
Hence, $G(4)\simeq-0.42$ and we again return to the above result
for the discord.
On the other hand, according to Eq.~(\ref{eq:cmT2}) the specific
heat at 4~K is $c_m(4)/R=0.4125$.
Solving the transcendental equation (\ref{eq:cmG}) and taking into
account that the temperature 4~K is larger than the maximum
temperature $T_{max}=1.82$~K, we find $G\simeq-0.4$ (see
Fig.~\ref{fig:cG}).
Then $Q\simeq0.18$. 

\subsection{Magnetic susceptibility}
\label{sec:sus}

Molar magnetic susceptibility of Heisenberg dimers satisfies
the Bleaney-Bowers equation \cite{BB52,Car86}
\begin{equation}
   \label{eq:chiBB}
   \chi(T)=\frac{N_A\,g^2\mu_B^2}{2k_BT}\,(1+G(T)) . 
\end{equation}
Here, $g$ is the corresponding component of the Land\'e factor when 
the measurements are made on a single crystal or
\begin{equation}
   \label{eq:g2}
   g^2={1\over3}\,(g_x^2 + g_y^2 + g_z^2)
\end{equation}
if the measurements are performed on a polycrystalline (powdered)
sample.

For the antiferromagnetic coupling $(J<0)$, the Bleaney-Bowers
susceptibility displays a maximum with coordinates
\begin{equation}
   \label{eq:Tmax-af}
   \frac{k_BT_{\rm max}}{|J|}=\frac{2}{1+W(3/e)}=1.2472\ldots ,
\end{equation}
\begin{equation}
   \label{eq:chi-max-af}
   \frac{|J|\chi_{\rm max}}{N_Ag^2\mu_B^2}={1\over3}\,W(3/e)
   =0.2011\ldots\, . 
\end{equation}
Here $W(x)$ is the Lambert function defined by the equation
$We^W=x$.
This function under the name LambertW($x$) was included in the
Maple package.

From Eq.~(\ref{eq:chiBB}), we get the spin correlation function
\begin{equation}
   \label{eq:Gchig}
   G(T)=\frac{2k_BT\chi(T)}{N_Ag^2\mu_B^2} - 1 .
\end{equation}
That is
\begin{equation}
   \label{eq:Gchi}
   G(T)=-1+{1\over2}\chi(T)/\chi_0(T) ,
\end{equation}
where
\begin{equation}
   \label{eq:chiC}
   \chi_0(T)=\frac{N_A\,g^2\mu_B^2}{4k_BT}
\end{equation}
is the Curie law for the paramagnetic ions.
Hence, one can indirectly measure the discord by performing
magnetometric measurements.

Note that the Bleany-Bowers susceptibility is related to the internal
energy (\ref{eq:u}) as
\begin{equation}
   \label{eq:uchi}
   u(T)=-3N_AJ\left(\frac{k_BT}{N_Ag^2\mu_B^2}\,\chi(T) -{1\over2}\right) .
\end{equation}
Consequently, the magnetic specific heat is
\begin{equation}
   \label{eq:cm}
   c_m(T)=-\frac{3k_BJ}{g^2\mu_B^2}\,\frac{\partial T\chi(T)}{\partial T} . 
\end{equation}
Moreover, inserting (\ref{eq:Gchig}) into equation (\ref{eq:cmG}) we
obtain an expression for the specific heat $c_m/R$ trough the magnetic
susceptibility without differentiation. 

We analyze, in the third instance, the sample with the copper nitrate
compound.
The magnetic susceptibility of ${\rm Cu(NO_3)_2\cdot2.5 H_2O}$
at low temperatures has been measured in Ref.~\cite{BFS63}.
The susceptibility of the powder has a rounded maximum at 3.2~K
where it is equal to $2\times0.065$~emu/mol.
For the model of binary clusters and isotropic exchange, the authors
\cite{BFS63} found from the data that $-J/2k_B=1.28$~K and $g=2.11$.
Using the experimental points presented in their figures
(see Ref.~\cite{BFS63}) and taking into account our normalizations per
a mole of dimers,
we found that at $T=4$~K the powder magnetic
susceptibility is $\chi=2\times0.063=0.126$~emu/mol.
Then, Eq.~(\ref{eq:Gchig}) yields $G(4)=-0.396$.
As a result, the discord is $Q=0.17$.
This estimate agrees more or less with the values obtained above from
the neutron scattering data and heat capacity measurements.
Some discrepancy can be attributed to experimental errors and
especially to the fact that the copper nitrate is only quasi-dimeric.

We will now move on to other compounds, which belong to the pronounced
binuclear materials.

\subsubsection{Antiferromagnetic substances}
\label{sec:af}

Let us consider classical examples of spin-${1\over2}$ Heisenberg
dimeric materials --- crystals of copper(II) acetates,
${\rm [Cu(CH_3COO)_2\cdot H_2O]_2}$ (hydrated) and
${\rm [Cu(CH_3COO)_2]_2}$ (anhydrous).
Their magnetic susceptibility study has a long history
beginning in 1915 and continuing into the present day 
\cite{Guha51,FM56,GMM71,El00,KS07}.
Experimental results between 90 and 400~K are given in table form
in \cite{FM56}.
The data have been described by the Bleany-Bowers equation with the
fitted parameters $2J/k_B=-408$ K, $g$=2.13 for the copper(II) acetate
monohydrate and $2J/k_B=-432$ K, $g$=2.17 for the anhydrous
copper(II) acetate \cite{FM56}.

\begin{figure}[t]
\begin{center}
\epsfig{file=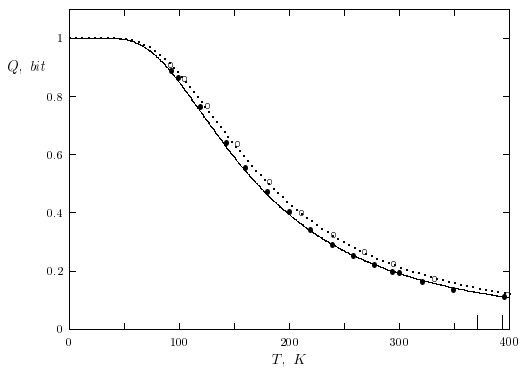,width=6.8cm}
\caption{
Thermal quantum discord in the hydrated $(-\!\!\bullet\!\!-)$ and
anhydrous $(-\!\circ\!-)$ copper(II) acetates.}
\label{fig:afCu}
\end{center}
\end{figure}

Using Eqs.~(\ref{eq:Q})--(\ref{eq:C}), (\ref{eq:Gchi}) we obtained
experimental points for the discord in both compounds.
Results are presented in Fig.~\ref{fig:afCu}.
Theoretical curves we plotted, taking the corresponding estimates
for the coupling constants $J/k_B$ and using Eqs.~(\ref{eq:G}),
(\ref{eq:Q})--(\ref{eq:C}).

We will now discuss the different domains in Fig.~\ref{fig:afCu}.
According to Eq.~(\ref{eq:QE}) the points which are higher than the
level $Q\simeq0.75$ belong to the region where entanglement
$E>Q$.
However, at temperatures $T>120-127$~K, the discord becomes larger
than the entanglement in both compounds.
The temperatures of $371$~K and $393$~K that are marked on the
abscissa axis by the longer bars are the points of sudden disappearance
of entanglement in hydrated and anhydrous copper(II) acetates,
respectively.
At higher temperatures, the entanglement is zero.
Near $400$~K, the discord is $11-12\%$ of the maximum
value $Q(0)=1$.

\subsubsection{Ferromagnetic compound}
\label{sec:f}
 
Entanglement in the ferromagnetic Heisenberg dimer is absent, 
but the discord is not zero.

We found in the recent literature the experimental data
for the ferromagnetic dimer material --- the binuclear copper(II)
acetate complex ${\rm [Cu_2L(OAc)]\cdot6H_2O}$, where
H$_3$L$=$2-(2-hydroxyphenyl)-1,3-bis[4-(2-hydroxyphenyl)-3-azabut-3-enyl]-1,3- imidazolidine
\cite{Fondo03}.
Magnetic susceptibility measurements for it have been performed
between 5 and $300$~K.
Taking into account the crystal structure of this compound, the
susceptibility data have been fitted to the Bleany-Bowers equation.
The best least-squares fit has been obtained with the parameters
$J/k_B=35.4$ K and $g=2.13$
\cite{Fondo03}.
At lowest measured temperatures near $5$~K, the experimental points
drop out from the theoretical dependence.
This may be ascribed to the influence of weak interdimer couplings.

\begin{figure}[t]
\begin{center}
\epsfig{file=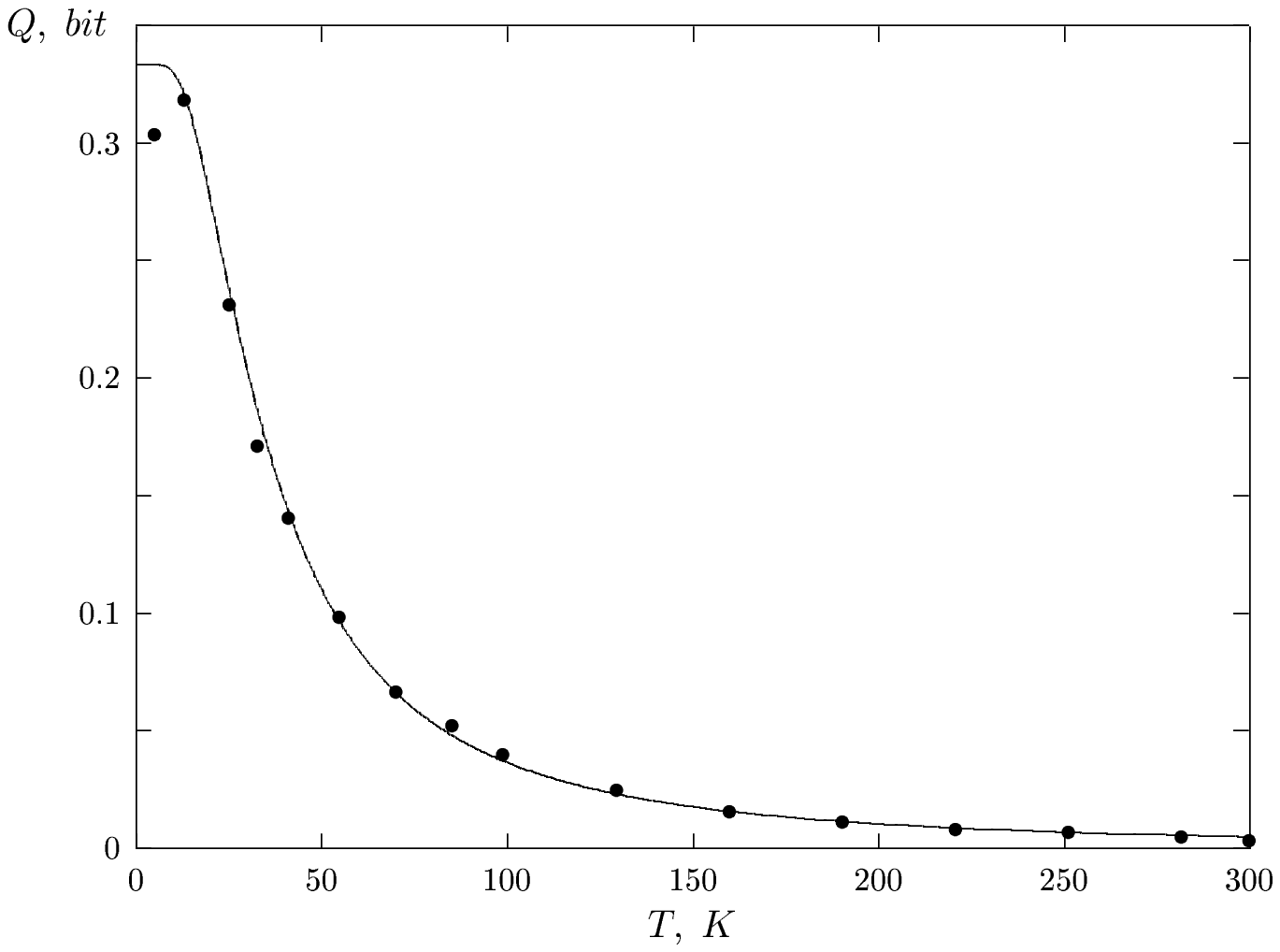,width=6.8cm}
\caption{
Quantum discord against the temperature in the ferromagnetic compound
${\rm [Cu_2L(OAc)]\cdot6H_2O}$.}
\label{fig:fCu}
\end{center}
\end{figure}

Using Eqs.~(\ref{eq:G}), (\ref{eq:Q})--(\ref{eq:C}), and
(\ref{eq:Gchi}) we obtained from magnetic data the temperature
behavior for the discord shown in Fig.~\ref{fig:fCu}.
Maximal discord, $Q\approx0.32$, is achieved near $13$~K.
At the temperature $T=300$~K, the product $\chi T=0.89$~cm$^3$K/mol
\cite{Fondo03}.
This leads to the discord $Q\approx0.003$, i.e. about $1\%$
relative to the theoretical limit of 1/3. 

\section{Conclusions}
\label{sec:Concl}

We have related the quantum discord in binuclear spin clusters to the
scattering data, as well as to the basic responses of a system to
external perturbations --- thermal and magnetic.
This makes it possible to measure the discord in various magnetic
substances.

In this paper, we have done a comparison analysis for the temperature
behavior of discord, classical correlations, entanglement, and
spin-spin correlation function in the two-qubit Heisenberg systems.
We have shown that in the case of antiferromagnetic interactions,
entanglement can be both larger and smaller than the discord or
classical correlations.
Discord and classical correlations can be present when entanglement
is absent.
This is observed both in antiferro- and ferromagnetic dimers.
By $T\neq0$, the discord is always larger than the classical
correlations.

We have presented the temperature dependencies of discord for
different solid-state dimeric materials with antiferro- and
ferromagnetic couplings.

In the current literature, the transfer of quantum correlations from
one system to another is intensively discussed (e.g., 
\cite{CBFPV06,Togan10,GT10} and references therein).
In particular, several schemes for extracting the entanglement from
solids have been reported.
With respect to discord, its dynamical behavior under the quantum state
transfer, for example, between semiconductor double-dot molecules
and photons is being studied now \cite{Pei10}.


\section*{ACKNOWLEDGMENT}
I am grateful to E.~I.~Kuznetsova for her help in the work.

This research was partially supported by the programs Nos.~18 and 21 of
the Presidium of RAS.



\end{document}